\documentclass[floatfix,aps,twocolumn,prb,superscriptaddress, ]{revtex4}
\usepackage{amssymb}
\usepackage{graphicx}
\usepackage{dcolumn}
\usepackage{natbib}
\usepackage{bm}

\input{tcilatex}

\begin{document}

\title{Infrared probe of the anomalous magnetotransport of highly oriented
pyrolytic graphite in the extreme quantum limit}
\author{Z.Q. Li}
\email{zhiqiang@physics.ucsd.edu}
\affiliation{Department of Physics, University of California, San Diego, La Jolla,
California 92093, USA}
\author{S.-W.~Tsai}
\affiliation{Department of Physics, University of California, Riverside, California
92521, USA}
\author{W.J.~Padilla}
\altaffiliation[Present address: ]{Department of Physics, Boston College, Chestnut Hill, MA 02467, USA}
\affiliation{Department of Physics, University of California, San Diego, La Jolla,
California 92093, USA}
\author{S.V.~Dordevic}
\affiliation{Department of Physics, The University of Akron, Akron, Ohio 44325, USA }
\author{K.S.~Burch}
\altaffiliation[Present address: ]{Los Alamos National Laboratory, MS K771, MPA-CIN, Los Alamos, NM 87545, USA}
\affiliation{Department of Physics, University of California, San Diego, La Jolla,
California 92093, USA}
\author{Y.J.~Wang}
\affiliation{National High Magnetic Field Laboratory, Tallahassee, Florida 32310, USA}
\author{D.N.~Basov}
\affiliation{Department of Physics, University of California, San Diego, La Jolla,
California 92093, USA}
\date{\today }

\begin{abstract}
We present\ a systematic investigation of the magneto-reflectance $R(\omega
,B)$ of highly oriented pyrolytic graphite (HOPG) in magnetic field $B$ up
to 18T. From these measurements, we report the determination of lifetimes $%
\tau $ associated with the lowest Landau levels in the quantum limit. We
find a linear field dependence for inverse lifetime 1/$\tau $($B$) of the
lowest Landau levels, which is consistent with the hypothesis of a
three-dimensional (3D) to 1D crossover in an anisotropic 3D metal in the
quantum limit. This enigmatic result uncovers the origin of the anomalous
linear in-plane magneto-resistance observed both in bulk graphite and
recently in mesoscopic graphite samples.
\end{abstract}

\maketitle

I. INTRODUCTION

\bigskip

Intense magnetic fields can radically alter properties of materials and
produce a myriad of novel effects and novel states of matter. \cite{Ando2D}%
\cite{DasSarma}\cite{Basov} One intriguing example is the field-induced
transformation expected in an anisotropic three-dimensional (3D) metal.
Provided only the lowest Landau levels (LLs) are populated (a condition
referred to as the quantum limit), coherent charge motion in the direction
perpendicular to the magnetic field is arrested. Therefore magnetic field
initiates a crossover from 3D to 1D transport along the direction of the
field.\cite{Abrikosov} A suitable candidate system to exhibit this
phenomenon is graphite. In this anisotropic semimetal comprised of weakly
coupled graphene sheets, a combination of low carrier density and small
effective masses allows one to achieve the quantum limit condition in fairly
modest magnetic fields along the c-axis (7-8T).\cite{review} A manifestation
of the 3D to 1D crossover is the field dependence of the relaxation rate 1/$%
\tau $ associated with the lowest LLs.\cite{Abrikosov} This fundamental
electronic characteristic is difficult to infer from transport studies but
can be directly probed via infrared (IR) spectroscopy.

Here, via IR spectroscopy of graphite we probe lifetimes $\tau $ associated
with the lowest LLs in the quantum limit with field perpendicular to the
graphene planes. We find that 1/$\tau \propto B$ for the lowest LLs, which
is distinct from power laws typically found in other classes of low
dimensional systems\cite{Ando2D} but consistent with the hypothesis of a 3D
to 1D crossover.\cite{Abrikosov} Empowered by the experimental data for
lifetimes, we were able to elucidate the anomalous linear magneto-resistance
in samples of weakly interacting graphene sheets. This peculiar form of
magneto-resistance was discovered nearly 40 years ago in bulk graphite\cite%
{McClure}\cite{review} and recently observed in mesoscopic samples comprised
of only few graphene layers.\cite{Zhang05}

\bigskip

II. EXPERIMENTAL METHODS

\bigskip

HOPG samples studied in this work were extensively characterized through
magneto-transport measurements.\cite{esq} They have ab-plane surfaces with
typical dimensions 6x6 mm$^{2}$. The zero field reflectance $R(\omega )$ was
measured over broad frequency (15 to 25,000 cm$^{-1}$) and temperature (10
to 292 K) ranges at UCSD. The complex optical conductivity $\sigma (\omega
)=\sigma _{1}(\omega )+i\sigma _{2}(\omega )$ in zero field was obtained
from Kramers-Kronig (KK) transformation of the reflectance data. The
Hagen-Rubens formula was employed in the low $\omega $ region. KK-derived
results are consistent with those obtained by ellipsometry.

The magneto-reflectance $R(\omega ,B)$ were measured in the Faraday
geometry: \textbf{E} vector in the graphene plane and magnetic field\ 
\textbf{B} parallel to the c axis of the sample. Our in-house apparatus
enables absolute measurements of $R(\omega ,B)$ in fields up to 9T over a
broad frequency range (15 to\ 4,000 cm$^{-1}$).\cite{Willie} Data in
selected magnetic fields were acquired using a sub-THz Martin-Puplett
interferometer and extend down to 7 cm$^{-1}.$ The magneto-reflectance
ratios $R(\omega ,B)/R(\omega ,B=0)$\ in the range 20-3,000 cm$^{-1}$ with
B=7-18 T were obtained at the National High Magnetic Field Laboratory in
Tallahassee. All data in magnetic field were taken at 5K.

\bigskip

III. ZERO FIELD\ DATA

\bigskip

The zero field reflectance $R(\omega )$ and dissipative part of the optical
conductivity $\sigma _{1}(\omega )$ are plotted in Fig. 1. The room
temperature $R(\omega )$ spectrum is consistent with earlier studies in the
overlapping frequency range.\cite{greenway} The $R(\omega )$ spectrum shows
a typical metallic behavior: a gradual depression of reflectance with
increasing frequency towards a plasma minimum at about 24,000 cm$^{-1},$
followed by a sharp onset due to the interband transition between $\pi $
bands at the M points of the Brillouin zone.\cite{review} An edge-like
structure in the far-IR reflectance shows an interesting temperature
dependence with softening of the edge from 400 cm$^{-1}$ at 292 K down to
200 cm$^{-1}$ at 10 K. The zero field $\sigma _{1}(\omega )$ spectra are
dominated by the Drude component but in addition show a substantial
contribution throughout the mid-IR range. This is detailed in the bottom
inset of Fig. 1 displaying a broad resonance corresponding to the edge
observed in raw $R(\omega )$ data. As temperature is reduced the Drude mode
narrowed indicative of strong depression of the quasiparticle scattering
rate $1/\tau $. We also found that mid-IR resonance softens with the center
peak shifting from 1,000 cm$^{-1}$ at 292 K down to 400 cm$^{-1}$at 10 K.

We assign the mid-IR resonance to the interband transitions between the $\pi
\ $bands near K and H points of the Brillouin zone,\cite{review} as shown in
Fig.~2. The conventional Slonzcewski-Weiss-McClure (SWMC) band theory of
graphite\cite{review} prescribes the following relations for the onsets of
interband transitions near K and H points:\cite{interband} $\Delta
E_{K}\lesssim 2(E_{F}-2\gamma _{2})$ and $\Delta E_{H}=\Delta -2E_{F},$
where $\Delta $ and $\gamma _{2}$ are the band parameters. Therefore, the
lowering of the Fermi energy\cite{review}\cite{EfT} and $\pi \ $bands near
the K point\cite{review} with the increase of T is expected to cause
hardening of the transitions at K and H points, as shown in Fig.~2. The
observed trends are consistent with the depression of $E_{F}$ by 40\% at
292K compared to the value at 10 K assuming accepted values for the Fermi
energy and the band parameters (at 10K): $E_{F}=-25meV,$ $\Delta
=0.04\thicksim 0.05eV$ \cite{delta} and $\gamma _{2}=-0.02eV$. Although this
effect has not been directly observed before, the evolution of the resonance
structure in $\sigma _{1}(\omega )$ with T is consistent with the well
established notion that the dominant contribution to the carrier dynamics
and electrodynamics of graphite is associated with K and H points of the
Brillouin zone.\cite{review} Therefore, our zero field results are
consistent with the SWMC band model. The applicability of SWMC model was
corroborated by recent angle resolved photoemission spectroscopy studies of
HOPG.\cite{Lanzara}

\bigskip

IV. MAGNETO-REFLECTANCE\ DATA

\bigskip

Insights into the anomalous magneto-transport can be provided by a
systematic investigation of magneto-reflectance $R(\omega ,B)$ spectra,
which are displayed in a $\omega /B$ scaled plot in Fig.~3. To highlight
field-induced changes of the optical properties, at each field $B$ the zero
field reflectance spectrum $R(\omega ,B=0)$ is also scaled by the same
factor $B$ and shown together with $R(\omega ,B)$ for comparison. Two new
features are observed in the high-field spectra: i) the reflectance at low
frequency is strongly suppressed and no longer extrapolates to unity in the
limit of $\omega \rightarrow 0$; ii) the $R(\omega ,B)$ spectra reveal a
series of field-dependent resonances extending up to 3,000 cm$^{-1}$ (see
also the bottom panel of Fig. 4). The dramatic suppression of the low
frequency reflectance $R(\omega \rightarrow 0,B)$ is an IR counterpart of
the enormous positive magnetoresistance of graphite exceeding 10$^{3}$ in 18
T field.\cite{review} The resonance features in $R(\omega ,B)$ originate
from the inter-LL transitions. An inspection of Fig. 3 shows that the
frequencies of the strongest resonances reveal a linear dependence on
magnetic field.

The rather complicated form of the $R(\omega ,B)$ spectra and its systematic
evolution with magnetic field are both in accord with the SWMC band theory
of HOPG. To show this we analyzed $R(\omega ,B)$ spectra displayed in Fig.~3
using a magneto-optical Lorentzian model for the complex conductivity due to
inter-LL transitions $\sigma _{\pm }^{LL}(\omega )$:\cite{Lax} 
\begin{equation}
\sigma _{\pm }^{LL}(\omega )=\sum_{j}\frac{\omega _{pj\pm }^{2}}{4\pi }\frac{%
\omega }{\omega /\tau _{j\pm }+i(\omega _{j\pm }^{2}-\omega ^{2})}
\end{equation}%
This equation yields a series of resonance peaks at $\omega _{j+}$ and $%
\omega _{j-}$\ at the poles of the left- and right-hand circularly polarized
conductivity $\sigma _{\pm }^{LL}(\omega )$. Each peak is defined through
the linewidth $1/\tau _{j\pm }$\ and the oscillator strength $\omega _{pj\pm
}^{2}$. Inter-LL transitions at both K and H points of the Brillouin zone%
\cite{Nakao}\cite{Toy} are included in our fitting. According to SWMC
theory, more than a dozen inter-LL transitions are likely to contribute to $%
R(\omega ,B)$ in the studied frequency range. However, the parameters in Eq.
(1) are strictly constrained by both the analytical properties of $\sigma
_{\pm }(\omega )$\cite{Lax} and the selection rules of inter-LL transitions.%
\cite{review}\cite{Toy}\ Following these constraints, Eq. (1) is used to
construct $\sigma _{+}(\omega )$ and $\sigma _{-}(\omega ),$ which are then
used to evaluate the reflectance spectra $R(\omega )=\frac{1}{2}%
[R_{+}(\omega )+R_{-}(\omega )]$. In addition, several field-independent
Lorentzian oscillators are employed in both $\sigma _{1+}(\omega )$ and $%
\sigma _{1-}(\omega )$ to account for the absorption above 3,000 cm$^{-1}$,
as displayed by the green (light gray) curve in Fig. 4. We observed an
additional resonance feature in $R(\omega ,B)$ below 8.5T in the range
50-100 cm$^{-1}/$T in Fig. 3, which is an expected consequence of the
quantum limit.\cite{Qlimit} The resonances in $R(\omega ,B)$ below and above
8.5T are roughly at the same positions in the $\omega /B$ plot, with small
deviations in the high frequency resonances. The two different sets of
magneto-optical oscillators at $\omega _{j\pm }/B$ employed to fit
experimental $R(\omega ,B)$ are summarized in table 1. We use the transition
energies from ref \cite{Toy} to account for the inter-LL transitions at the
H point. These transitions are at much higher energy compared to those at
the K point and constitute a broad background in $R(\omega ,B)$ spectra.\cite%
{Toy} We stress that all field-dependent oscillators employed in our model
can be assigned to the inter-LL transitions at K and H points within the
conventional SWMC band model.\cite{review}

The model spectra at representative fields obtained from the above analysis
are depicted in Fig. 3 and 4. Both the peak structures and their field
dependence in experimental $R(\omega ,B)$ spectra are reproduced by our
analysis. Of special interest is the lowest inter-LL transition (LL-1)-(LL0)
shown in Fig. 5. This transition is well separated from higher energy
resonances, therefore the reflectance in the vicinity of this mode is hardly
affected by other higher energy modes in our model, which allows us to
extract the parameters for this particular transition from the above
analysis. Moreover, in the quantum limit LL-1 and LL0 are the only occupied
levels,\cite{review} which are directly related to the magneto-transport
properties. Therefore, special attention will be given to the (LL-1)-(LL0)
transition. Both the resonance frequency of this transition and its
linewidth 1/$\tau $ follow a linear field dependence as shown in Fig. 5.
This is evident from a close inspection of the experimental $R(\omega ,B)$
spectra in Fig.~3. The power law of 1/$\tau (B)$ below 6T field is yet to be
explored, because at low fields the (LL-1)-(LL0) transition occurs at low
energies beyond our detection limit.

\bigskip \bigskip

V. DISCUSSION

\bigskip

Our magneto-reflectance data are in accord with the conventional SWMC band
model of graphite.\cite{review} In the SWMC model, the inter-LL transitions
at K and H points have different field dependence: transitions at the K
point reveal \textquotedblleft normal\textquotedblright\ quasiparticle
behavior (linear $B$ scaling),\cite{review} whereas transitions at the H
point show Dirac behavior at high fields ($\sim $ $\sqrt{B}$ scaling).\cite%
{Toy} Inter-LL transitions at both K and H points of the the Brillouin zone
are needed to fully account for our results on HOPG, which is consistent
with previous magneto-optical study of bulk graphite samples.\cite{Toy}
Recently, inter-LL transitions with $\sqrt{B}$ scaling were observed in
magneto-transmission study of ultrathin graphite layers,\cite{deHeer} which
may come from transitions at the H point due to inter-layer coupling between
graphene sheets.

We now show that the observed 1/$\tau \propto B$ dependence for the
(LL-1)-(LL0) transition (Fig. 5) is consistent with the notion of 3D to 1D
crossover in an anisotropic 3D metal in the quantum limit.\cite{Abrikosov}
The 1/$\tau \propto B$ behavior in Fig. 5 most likely originates from the
linear field dependence of the broadening of both LL-1 and LL0,\cite{tau}
which defines the scattering rate of the quasiparticles associated with
these levels. In the quantum limit, the carriers have finite momentum only
along the field direction (c-axis) and move diffusively in the transverse
direction (ab plane) when scattered by impurities. The carrier-impurity
scattering rate 1/$\tau $ in the quantum limit is given by 1/$\tau (B)=2\pi
n_{0}u_{0}^{2}\nu _{B}$,\cite{Abrikosov} where the density of neutral
impurities $n_{0}$ and impurity potential $u_{0}$ are $B$-independent. The
3D density of states $\nu _{B}$ increases linearly with $B$ in the quantum
limit in graphite,\cite{review} giving rise to the linear $B$ dependence of
1/$\tau .$ The increase of 1/$\tau $ by a factor of 3 from 6T to 18T is
consistent with a threefold increase of $\nu _{B}$ in this range.\cite%
{review} The interactions between carriers can give additional corrections
to 1/$\tau (B),$ which have weak logarithmic field dependence.\cite%
{Shan-Wen1} Therefore our data for the field dependence of the relaxation
rate are in accord with the specific prediction of a model involving 3D to
1D crossover in the quantum limit.

Based on the observed 1/$\tau \propto B$ result, we propose an explanation
for the anomalous linear magneto-resistance found both in mesoscopic
graphite systems\cite{Zhang05} and in bulk graphite.\cite{review} First, the
giant positive magneto-resistance of graphite is clearly related to the
electronic spectral weight transfer of the Drude mode in the $B$=0 spectrum
to finite energies in magnetic field (Fig. 4). The spectral weight transfer
results in a drastic suppression of the conductivity in the $\omega
\rightarrow 0$ limit by more than 10$^{3}$. Data in Figs. 3, 4 furthermore
show that the lowest energy electronic excitation in the quantum limit is
the (LL-1)-(LL0) transition revealing the linear field dependence of the
relaxation rate. Let us explore the consequences of these findings for
magneto-resistance that we will analyze following the standard form
appropriate for the quantum limit as shown in:\cite{Mermin}\cite{Abrikosov}

\begin{equation}
\rho _{xx}=\frac{\rho _{1}\rho _{2}(\rho _{1}+\rho _{2})+(\rho
_{1}R_{2}^{2}+\rho _{2}R_{1}^{2})B^{2}}{(\rho _{1}+\rho
_{2})^{2}+(R_{1}+R_{2})^{2}B^{2}}
\end{equation}%
\bigskip

where $\rho _{xx}$ is the ab-plane resistivity, $\rho _{i}$ and R$_{i}=$1/n$%
_{i}$q$_{i}$ are the resistivity and Hall coefficient of the electrons (i=1)
and holes (i=2), and n$_{i}$ and q$_{i}$ are the density and charge of the
carriers. Note that in the quantum limit the off-diagonal term in the
resistivity tensor can still be expressed as R$_{i}B$ as shown in Ref.\cite%
{Abrikosov}. We emphasize that here Eq. (2) follows from a fully quantum
mechanical treatment yielding the magnetic field dependence of $\rho _{i}$
and R$_{i}$. This is distinct from the semiclassical model of
magnetoreistance, in which $\rho _{i}$ and R$_{i}$ are constants. At high
fields, the $B^{2}$ term dominates over the zero field term in the
numerator, therefore the latter can be neglected. In the quantum limit,
transport in the ab plane is diffusive with $\rho _{i}$ given by 1/$\rho _{i}
$ $\sim $ e$^{2}\nu _{B}$D,\cite{Shan-Wen1} where D $\sim $ $l_{B}^{2}/\tau $
is the diffusion coefficient and $l_{B}=1/\sqrt{eB}$ is the magnetic length.
From $\nu _{B}$ $\propto B,$\cite{review} and D $\sim $ $l_{B}^{2}/\tau $ $%
\propto B$/$B$ $\sim $1 (1$/\tau \propto B$ from our results), we obtain 1/$%
\rho _{i}\propto B$. Because the net carrier density in the sample $%
n_{1}-n_{2}$ is constant at high field\cite{review}\cite{Brandt} and $%
n_{1}\approx n_{2}\propto B$, we obtain $(R_{1}+R_{2})B=\frac{(n_{1}-n_{2})}{%
n_{1}n_{2}e}B\propto 1/B.$ This expression is valid in the magnetic field
regime where graphite shows linear magneto-resistance. In even higher field, 
$n_{1}-n_{2}$ decreases with field due to the so-called magnetic freeze-out
effect, and correspondingly a saturation of magneto-resistance is observed.%
\cite{Brandt} Combined with $R_{i}=1/n_{i}e\propto 1/B$, we find $\rho
_{xx}\propto B$. Therefore, our study suggests that the anomalous linear
magneto-resistance of graphite originates from quantum transport in magnetic
field. The above analysis can also be used to account for the linear
magneto-resistance in mesoscopic graphite-based electronic systems,\cite%
{Zhang05} where interlayer coupling between graphene sheets is important.

The use of a two-band model in our analysis is corroborated by a recent
study of mesoscopic graphite field-effect transistors,\cite{Zhang05} which
shows that the slope of the magneto-resistance can be tuned by changing the
carrier compensation $n_{e}-n_{h}$ consistent with Eq (2). Previously, a
model based on magnetic-field-dependent scattering range of ionized impurity
scattering\cite{McClure} was proposed to explain the linear
magneto-resistance in graphite. Due to the field-dependent screening of
charged impurities, the scattering rate in this model has a field dependence
of 1$/\tau \sim B/(c+B)^{2}$, where c is a constant.\cite{McClure} This
prediction is in contrast to our observation 1$/\tau \propto B$. Therefore,
our study shows that impurities in graphite are likely to be uncharged and
the ionized impurity scattering model\cite{McClure} may not be applicable.
Indeed, previous magneto-transport studies have shown the limitations of the
ionized impurity model.\cite{Dresselhaus1}\cite{Dresselhaus2} Recently, an
interesting model based on Dirac fermions\cite{AbriGraphite} is proposed to
account for linear magneto-resistance in graphite. However, our studies of
HOPG have clearly demonstrated that Dirac fermions do not play a dominant
role in the magneto-transport in graphite.

\bigskip

VI. SUMMARY

\bigskip

In this study, we carried out systematic IR magneto-optical experiments on
HOPG. We show that the optical properties of HOPG in both zero field and
magnetic fields can be understood within the conventional SWMC band theory
of graphite.\cite{review} We observed an unconventional linear field
dependence of the inverse lifetime 1/$\tau $ associated with the lowest LLs.
This observation is consistent with the hypothesis of 3D to 1D crossover in
an anisotropic 3D metal in the quantum limit.\cite{Abrikosov} This new
result has allowed us to address the origin of the anomalous linear
magneto-resistance in mesoscopic graphite-based electronic systems\cite%
{Zhang05} as well as in bulk graphite.\cite{review} Our work has
demonstrated the potential of IR spectroscopy for the study of LL\ dynamics
and novel magneto-transport phenomena in carbon based materials such as
mesoscopic graphite and graphene,\cite{CastroNeto}\cite{sharapov} which are
directly related to the novel functionalities of magneto-electronic devices
based on these materials.

\bigskip

ACKNOWLEDGMENTS

\bigskip

This work was supported by the DOE and NSF. We thank P. Esquinazi for
providing samples.

\begin{table*}[tbp]
\caption{The magneto-optical Lorentzian oscillators used in the analysis of R%
$(\protect\omega ,B)$ and their possible assignments. $\protect\omega _{+}/B$
and $\protect\omega _{-}/B$\ (in cm$^{-1}/T$) are frequencies (normalized by
magnetic field $B$) of the oscillators in $\protect\sigma _{+}(\protect%
\omega )$ and $\protect\sigma _{-}(\protect\omega )$, repectively, used to
account for the inter-Landau-level (LL) transitions at K-point. Two sets of
oscillators are used in two field ranges: (\textit{h}) $B>$8.5T and (\textit{%
l}) $B\leqq $8.5T. The theoretical transition frequencies are labeled
following Nakao's notation\protect\cite{Nakao}: hn-em refers to the
transition from the n-th hole LL to the m-th electron LL, etc. The
transitions marked with stars are allowed only below the quantum limit. The
last 8 columns list $\protect\omega _{pj\pm }$ and $1/\protect\tau _{j\pm }$
(both in cm$^{-1}$) used for the model spectrum R($\protect\omega $) at\ 16
and 8.5T. The parameters of the oscillator used for m$^{+}$=0,-1 transitions
at H-point are: 16 T: $\protect\omega _{0}=1212$ cm$^{-1},\protect\omega %
_{p}=4243$cm$^{-1},1/\protect\tau =500$cm$^{-1};$ 8.5T: $\protect\omega %
_{0}=818$ cm$^{-1},\protect\omega _{p}=3211$cm$^{-1},1/\protect\tau =300$cm$%
^{-1}.$}%
\begin{ruledtabular}

\begin{tabular}{cccccccccccccccccccc}
$\omega _{-}^{h}$/B&
$\omega _{-}^{l}$/B&
theory&
assignment&
$\omega _{+}^{h}$/B&
$\omega _{+}^{l}$/B&
theory&
assignment&
$\omega _{pj-}^{16T}$&
$\omega _{pj+}^{16T}$&
1/$\tau _{j-}^{16T}$&
1/$\tau _{j+}^{16T}$&
$\omega _{pj-}^{8.5T}$&
$\omega _{pj+}^{8.5T}$&
1/$\tau _{j-}^{8.5T}$&
1/$\tau _{j+}^{8.5T}$&
\tabularnewline
\hline
\hline 
4.6 & 4.3 & 4.5 & (LL-1)-(LL0) & 19.7 & 19.3 & 22 & (LL-1)-e1 & 1238 & 2265 & 10 & 40 & 725 & 1347 & 6 & 24 \\ 
13 & 18 & 17 & (LL0)-e1 & 28 & 33 & 35 & (LL0)-e2 & 4016 & 5578 & 29 & 60 & 3683 & 4953 & 5 & 10 \\ 
35 & 35 & 32 & h2-(LL0) & 50 & 50 & 49 & h2-e1 & 2079 & 2377 & 167 & 227 & 2325 & 2629 & 96 & 128 \\ 
\hline 
54 &  & 54 & h1-e2 & 69 &  & 70 & h1-e3 & 2082 & 2349 & 26 & 33 &  &  \\ 
& 58 & 58 & h3-e1 &  & 73 & 70 & (LL-1)-e4 &  &  &  &  & 1959 & 2130 & 7 & 9 \\ 
\hline 
70 & 61 & 66 & (LL0)-e4 & 85 & 76 & 76 & h3-e2 & 1538 & 1672 & 146 & 175 & 1581 & 1753 & 36 & 45 \\ 
&  &  &  &  &  & 81 & (LL0)-e5 &  &  &  &  &  \\ 
\hline 
83 & 80 & 84 & h2-e3 & 98 & 95 & 98 & h2-e4 & 1056 & 1143 & 67 & 78 & 2340 & 2519 & 92 & 108 \\ 
&  & 85 & h4-e2 &  &  & 101 & h4-e3 &  &  &  &  \\ 
\hline 
100 &  & 99 & h1-e5 & 115 &  & 113 & h1-e6 & 2461 & 2615 & 210 & 240 &  &  \\ 
& 111 & $\star $105 & $\star $e1-e8 &  & 126 & $\star $119 & $\star $e1-e9 & 
&  &  &  & 2158 & 2293 & 70 & 79 \\ 
&  & 107 & h3-e4 &  &  & 121 & h3-e5 &  &  &  &  \\ 
&  & 110 & h5-e3 &  &  & 124 & h5-e4 &  &  &  &  \\ 
\hline 
122 &  & 127 & h2-e6 & 137 &  & 141 & h2-e7 & 1874 & 1977 & 128 & 143 &  &  \\ 
&  & 130 & h4-e5 &  &  & 144 & h4-e6 &  &  &  &  \\ 
\hline 
& 141 & 141 & h1-e8 &  & 156 & 155 & h1-e9 &  &  &  &  & 2213 & 2324 & 73 & 80 \\ 
145 &  & 149 & h3-e7 & 160 &  & 163 & h3-e8 & 2305 & 2409 & 211 & 232 &  &  \\ 
&  & 153 & h5-e6 &  &  & 167 & h5-e7 &  &  &  &  \\ 
\hline 
& 164 & 169 & h2-e9 &  & 179 & 183 & h2-e10 &  &  &  &  & 2343 & 2442 & 101 & 110 \\ 
167 &  & 172 & h4-e8 & 182 &  & 186 & h4-e9 & 1498 & 1559 & 172 & 187 &  &  \\ 
&  & 175 & h6-e7 &  &  & 189 & h6-e8 &  &  &  &
\tabularnewline
\end{tabular}
\end{ruledtabular}
\end{table*}

\begin{figure}[tbp]
\caption{(Color online) Zero field ab-plane reflectance $R(\protect\omega )$
(top panel) and real part of the optical conductivity $\protect\sigma _{1}(%
\protect\omega )$ (bottom panel) of graphite. The thin dashed lines in the
bottom panel are Drude fits of the $\protect\sigma _{1}(\protect\omega )$
spectra in the low frequency region. The bottom inset displays the
absorption peak of $\protect\sigma _{1}(\protect\omega )$ in the mid-IR
region, which shifts to high energy with increasing temperature. }
\end{figure}

\begin{figure}[tbp]
\caption{(Color online) Schematics of the band structure of graphite at K
and H points of the Brillouin zone at 10K\ and 292K. The arrows indicate the
interband transtions observed in the bottom inset of Fig. 1.}
\end{figure}

\begin{figure}[tbp]
\caption{(Color online) The $\protect\omega /B$\ scaled plot of the ab-plane
reflectance spectra $R(\protect\omega ,B)$\ in several magnetic fields. At
each field $B$, the zero field reflectance spectrum $R(\protect\omega ,B=0)$
is also scaled by the corresponding factor $B$ for comparison. Every set of
spectra is offset by 30\% for clarity. The model spectra are obtained from
the analysis detailed in the text.}
\end{figure}

\begin{figure}[tbp]
\caption{(Color online) Top panel: $\protect\sigma _{1}(\protect\omega )$ at
zero field and model spectra $\protect\sigma _{1+}(\protect\omega )$ and $%
\protect\sigma _{1-}(\protect\omega )$ at 16T obtained from the analysis
detailed in the text$.$ The black square on the left axis represents the DC
conductivity $\protect\sigma _{DC}$ at 0T. The green (light gray) curve is
the high frequency field-independent absorption in both $\protect\sigma %
_{1-}(\protect\omega )$ and $\protect\sigma _{1+}(\protect\omega )$. Bottom
panel: experimental R($\protect\omega $) spectra at 0T and 16T and model
spectra R($\protect\omega $) at 16T. The arrows indicate the resonance
features in R($\protect\omega $) and $\protect\sigma _{1-}(\protect\omega )$%
\ due to the (LL-1)-(LL0) transition.}
\end{figure}

\begin{figure}[tbp]
\caption{(Color online) Resonance frequency (left panel) and linewidth
(middle panel) of the (LL-1)-(LL0) transition\ in magnetic field. The dashed
curves in these two panels are linear fits to the data. The dotted curves in
the left panel are $\protect\omega \sim \protect\sqrt{B}$ fits. Right panel:
a schematic of the LL-1 and LL0 Landau levels in the quantum limit (adapted
from Ref. \protect\cite{Dresselhaus2}).}
\end{figure}


\begin{thebibliography}{99}
\bibitem{interband} From the expressions and structures of the four $\pi $
bands in the SWMC model (see ref. \cite{review}), the onset energy of
interband transitions at K and H points can be calculated using the
particular values of parameters at these points.
\bibitem{Qlimit} The different resonance features below and in the quantum
limit can be attributed to the different allowed inter-LL transitions in
these two ranges. For instance, the inter-LL transitions such as e1-e8 and
e1-e9 are only allowed below the quantum limit, as detailed in Table 1.
\bibitem{tau} In 2D systems under strong magnetic field, the resonance
linewidth of the inter-LL transition coincides with the broadening of the
corresponding LLs, provided impurity scattering is the dominant relaxation
process.\cite{Ando2D}
\end{thebibliography}
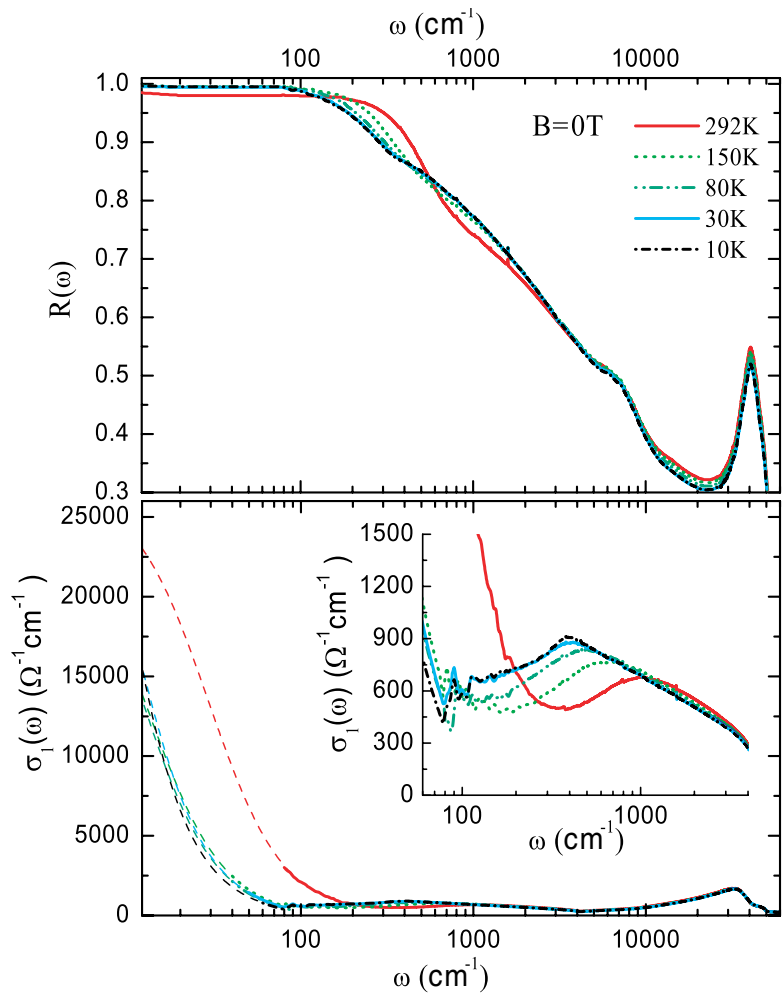
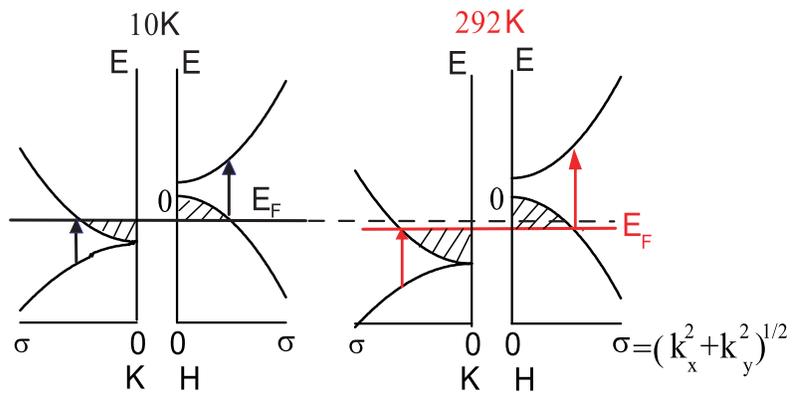
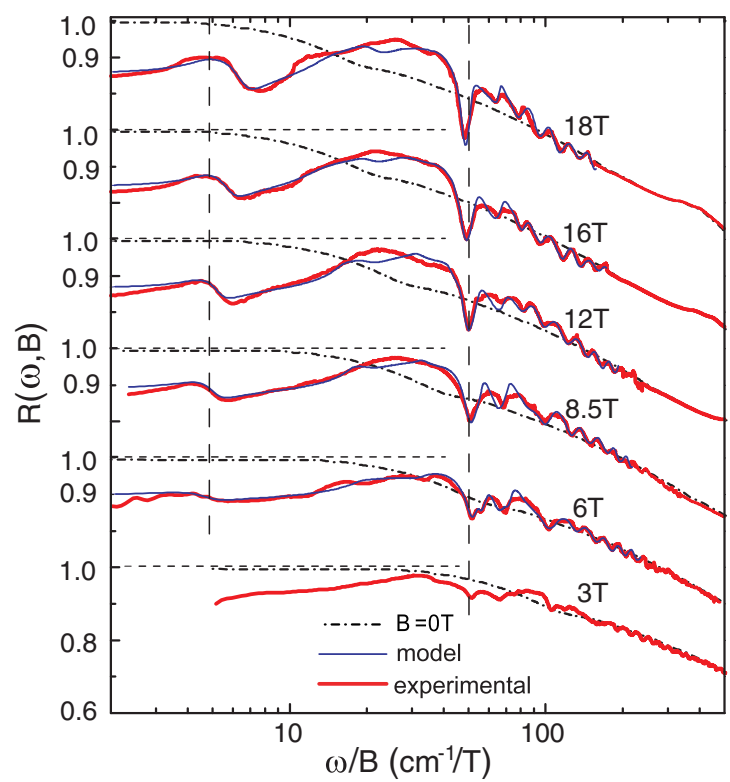
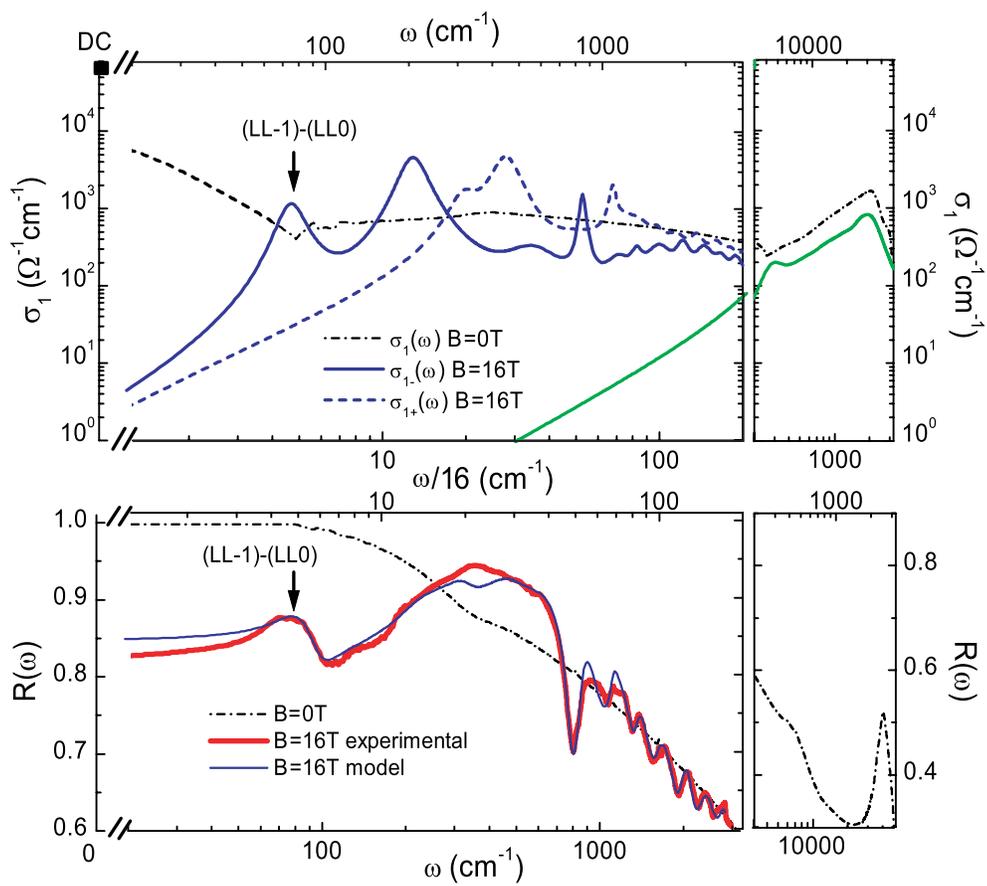
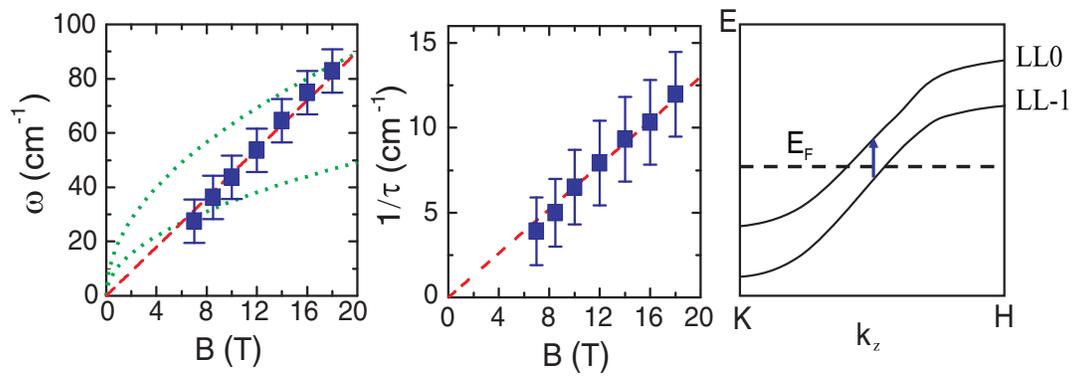
\end{document}